\documentclass[twocolumn,preprintnumbers,amsmath,aps]{revtex4}
\usepackage{graphicx}
\usepackage{dcolumn}
\usepackage{bm}
\usepackage{subfigure}
\usepackage[usenames]{color}
\begin{document}
\newcommand{\kvec}{\mbox{{\scriptsize {\bf k}}}}
\def\eq#1{(\ref{#1})}
\def\fig#1{\ref{#1}}
\def\tab#1{\ref{#1}}
\title{Influence of lithium doping on the thermodynamic properties of graphene based superconductors}
\author{D. Szcz{\c{e}}{\'s}niak$^{1}$}\email{d.szczesniak@ajd.czest.pl}
\author{A.P. Durajski$^{2}$}
\author{R. Szcz{\c{e}}{\'s}niak$^{2}$}
\affiliation{1. Institute of Physics, Jan D{\l}ugosz University in Cz{\c{e}}stochowa, 
Al. Armii Krajowej 13/15, 42-200 Cz{\c{e}}stochowa, Poland}
\affiliation{2. Institute of Physics, Cz{\c{e}}stochowa University of Technology, Al. Armii Krajowej 19, 42-200 Cz{\c{e}}stochowa, Poland}
\date{\today} 
\begin{abstract}
The superconducting phase in graphene can be induced by doping its surface with the lithium atoms. 
In the present paper, it has been shown that the critical temperature ($T_C$) for the LiC$_{6}$ and Li$_{2}$C$_{6}$ compounds change from $8.55$ K to $21.83$ K. The other thermodynamic parameters: the order parameter ($\Delta$), the specific heat for the superconducting ($C^{S}$) and the normal ($C^{N}$) state and the thermodynamic critical field ($H_{C}$) differ from the predictions of the BCS theory. In particular, the ratio 
$R_{\Delta}\equiv 2\Delta\left(0\right)/k_{B}T_{C}$ is equal to: 
$\left[3.72\right]_{{\rm LiC}_6}$ and $\left[4.21\right]_{{\rm Li}_{2}{\rm C}_6}$. Additionally, the quantities 
$R_{C}\equiv\Delta C\left(T_{C}\right)/C^{N}\left(T_{C}\right)$ and 
$R_{H}\equiv T_{C}C^{N}\left(T_{C}\right)/H_{C}^{2}\left(0\right)$ take the values:
$\left[1.47\right]_{{\rm LiC}_6}$, $\left[1.79\right]_{{\rm Li}_{2}{\rm C}_6}$, and 
$\left[0.167\right]_{{\rm LiC}_6}$, $\left[0.144\right]_{{\rm Li}_{2}{\rm C}_6}$. 
Finally, it has been shown that the electron effective mass at $T_C$ is high: 
$\left[1.61m_e\right]_{{\rm LiC}_6}$ and $\left[2.12m_e\right]_{{\rm Li}_{2}{\rm C}_6}$.
\end{abstract}
\maketitle

\section{Introduction}

At present, carbon allotropes constitute one of the most popular and promising research fields in the condensed matter physics \cite{hirsch}, \cite{szczesniak1}, \cite{yang}. In particular, the special attention is given to the two-dimensional one-atom-thick carbon structure known as graphene \cite{novoselov},  \cite{castro1}. The great interest in this material is driven by its numerous extraordinary electronic, thermal, and mechanical properties  \cite{castro2}, \cite{balandin}, \cite{lee}. As a consequence of these superior features, graphene is expected to have a variety of applications, particularly as a potential building block for the future electronic devices \cite{schwierz}, \cite{pasanen}. However, the usefulness of pristine graphene for the electronics is somehow limited due to its semimetallic character.

The semiconducting energy gap in graphene can be opened by the structural or chemical modifications of its pristine form \cite{shimizu}, \cite{fan}. The structural modifications or the chemical doping can also enable the induction of the superconducting phase. We notice that the existence of the superconducting state in graphene is important since it may allows to extend available range of carbon-based nanoelectronics towards more efficient superconductor-quantum dot devices \cite{franceschi} or low-dimensional superconducting transistors \cite{delahaye}.

Despite of the semi-metallic character of pristine graphene (the low density of states at the Fermi level), there are two other reasons that make difficult the induction of the superconducting state in this material. First, the in-plane vibrations in graphene are very energetic. Second, there is no coupling between the in-plane $\pi$-type states and the out-of-planes vibrations \cite{calandra}.

In order to overcome these issues few solutions to this problem have been proposed.

The pioneering works predominantly concentrated on the existence of the Dirac points in the electron band energy function \cite{uchoa}. Another attempts concerned shifting Fermi energy close to the van Hove singularity, in order to introduce mobile charge carriers above Dirac points \cite{nandkishore}. It should be noticed that this approach considered the unconventional pairing mechanism.

The preliminary investigations of the phonon-mediated superconducting state in graphene has been given in \cite{lozovik} and \cite{einenkel}. Similarly as in \cite{nandkishore}, the investigations of the electron-phonon superconducting state has been made for energies above the Dirac point, but this time not close to the van Hove singularity. In particular, these papers discussed the valley structure of the order parameter and only suggested the possiblity of inducing the superconducting state in graphene.

First quantitative predictions of the closely related graphene superconducting material known as graphane \cite{elias} has been proposed by Savini {\it et al.} \cite{savini} on the basis of the first-principles calculations. It has been shown, that the {\it p}-doped graphane may leads to the superconducting transition temperature of the value above the boiling point of liquid nitrogen.

Another progress in the research on the conventional superconductivity in graphene has been made recently, when it has been shown that the phonon-mediated superconducting state can be induced via deposition of the lithium atoms on its surface \cite{profeta}. In this study Profeta {\it et al.} recalled the investigations on the graphite intercalated compounds and showed that due to the removal of quantum confinement, the lithium adatoms in graphene generates additional intralayer states on the Fermi level giving rise to the reasonably high electron-phonon coupling constant.

Although these findings are still not experimentally confirmed, the recent experimental results present that the closely related lithium-intercalated bilayer graphene structures are stable \cite{sugawara}. Hence, the theoretical predictions given by Profeta {\it et al.} can be considered as the interesting and promising.

In the presented paper, we have supplemented the results obtained in \cite{profeta} by calculating the thermodynamic properties of LiC$_{6}$ and Li$_{2}$C$_{6}$ superconductors. Our analysis has been conducted in the framework of the Eliashberg formalism \cite{eliashberg}, \cite{carbotte}, which represents the strong-coupling generalization of the Bardeen-Cooper-Schrieffer theory (BCS) \cite{bardeen1}, \cite{bardeen2}.

\section{Theoretical model}

For the phonon-mediated superconductors, the thermodynamic properties can be derived from the knowledge of the Eliashberg spectral function $\alpha^2F(\Omega)$, where $\alpha$ denotes the average electron-phonon coupling, $F(\Omega)$ represents the phonon density of states, and $\Omega$ is the phonon frequency.

The $\alpha^2F(\Omega)$ function can be obtained within the first-principles methods or by the analysis of the tunnelling data. 
In the presented paper, we have taken into consideration the $\alpha^2F(\Omega)$ functions for LiC$_6$ and Li$_2$C$_6$ compounds, which have been calculated in \cite{profeta}. We notice that these functions have been obtained by using the first-principles methods (QUANTUM ESPRESSO package \cite{perdew}, \cite{giannozzi}).

The Eliashberg equations have been solved on the imaginary axis and in the mixed representation (simultaneously on the imaginary and real axis), using the iterative method presented in the papers \cite{szczesniak2}, \cite{szczesniak3}, and \cite{szczesniak4}. In our calculations, we have taken into consideration the $1100$ Matsubara frequencies: $\omega_{m}\equiv\frac{\pi}{\beta}(2m-1)$, where parameter $\beta$ is given by $\beta\equiv 1/k_{B}T$, and $k_{B}$ denotes the Boltzmann constant. Such assumption ensures the stability of the numerical solutions for $T\geq T_{0}\equiv 1.75$ K.

Finally, the Coulomb pseudopotential ($\mu^{\star}$), which models the depairing electron correlations is equal to $0.1$. The cut-off frequency ($\omega_{c}$) is set to be $\omega_{c}=3\Omega_{\rm max}$, where $\Omega_{\rm max}$ is the maximum phonon frequency equals to $195.25$ meV and to $187.48$ meV for LiC$_6$ and Li$_2$C$_6$ superconductors, respectively.

\section{Results and discussion}

We begin our analysis with the calculation of the order parameter on the imaginary axis ($\Delta_{m}\equiv\Delta\left(i\omega_{m}\right)$).

\begin{figure}[ht]
\includegraphics[width=\columnwidth]{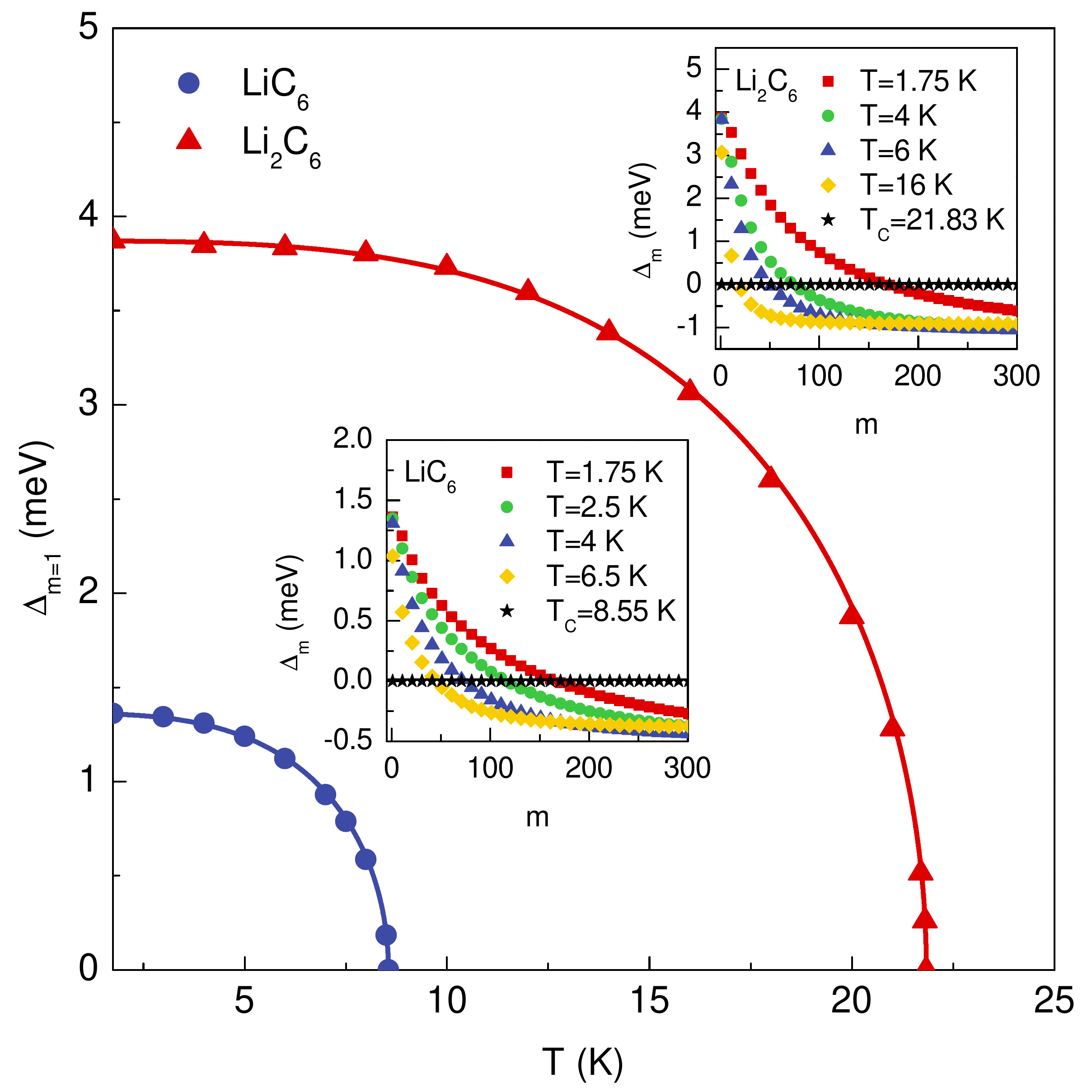}
\caption{The dependence of the maximum value of the order parameter on the temperature for LiC$_6$ and Li$_2$C$_6$ monolayers. 
Circles and triangles represent the exact Eliashberg solutions, whereas the solid lines have been obtained on the basis of the formula (\ref{r1}). The insets present the values of the order parameter as a function of $m$ for selected temperatures.}
\label{f1}
\end{figure}

In figure \ref{f1}, we have presented the dependence of the maximum value of the order parameter ($\Delta_{m=1}$) on temperature. The above results have been obtained on the basis of the data plotted in the insets, where the order parameter as a function of $m$ has been shown.

It can be observed that the value of $\Delta_{m=1}$ strongly decreases with the growth of the temperature and the decreasing lithium doping. This fact can be exactly parameterized by using the simple formula:
\begin{equation}
\label{r1}
\Delta_{m=1}=\Delta_{m=1}\left(0\right)\sqrt{1-\left(\frac{T}{T_{C}}\right)^{\Gamma}}, 
\end{equation}
where $\Delta_{m=1}\left(0\right)\equiv\Delta_{m=1}\left(T_{0}\right)$ is equal to $1.36$ meV and to $3.87$ meV for LiC$_6$ and Li$_2$C$_6$, respectively.
In both cases: $\Gamma=3.25$.

We notice that, in the first approximation, the $2\Delta_{m=1}$ function allows us to determine the value of the energy gap at the Fermi level.

The value of the critical temperature ($T_{C}$) has been extracted on the basis of the equation: $\left[\Delta_{m=1}\right]_{T_{C}}=0$. 
The obtained values of $T_{C}$ are equal to $8.55$ K and to $21.83$ K for LiC$_6$ and Li$_2$C$_6$ superconductor.

\begin{figure}[ht]
\includegraphics[width=\columnwidth]{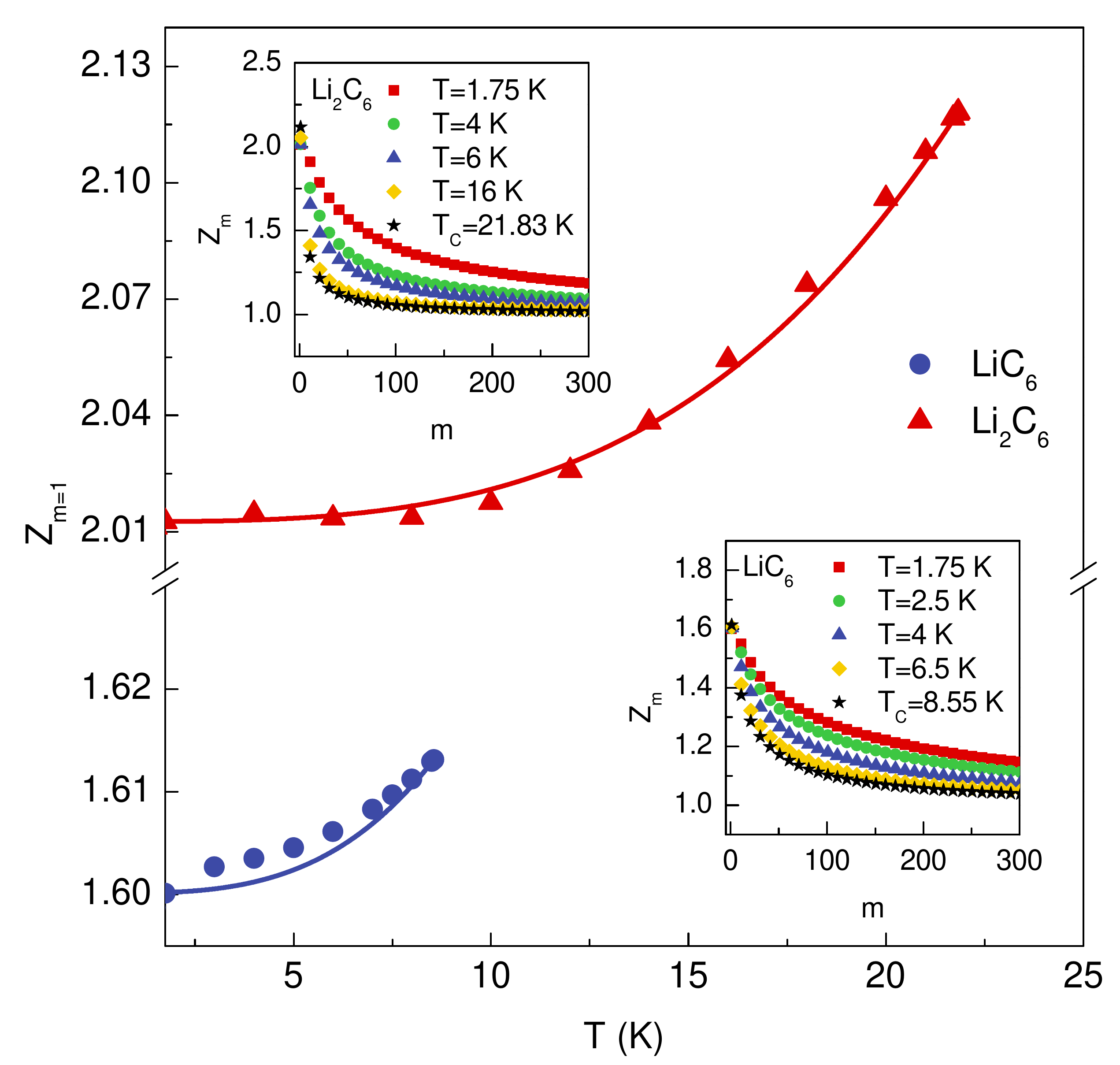}
\caption{The dependence of the maximum value of the wave function renormalization factor on the temperature for  LiC$_6$ and Li$_2$C$_6$. Circles and triangles represent the exact Eliashberg solutions, whereas the solid lines are obtained on the basis of the analytical formula (\ref{r2}). 
The insets present the values of the wave function renormalization factor as a function of $m$ for selected temperatures.}
\label{f2}
\end{figure}

In contrast to the order parameter, the maximum value of the wave function renormalization factor ($Z_{m=1}$) grows with increasing temperature, as presented in figure \ref{f2}. However, $Z_{m=1}$ increases also rapidly with the lithium doping.

Similarly as in the case of the order parameter, the dependence of $Z_{m=1}$ on the temperature for considered lithium doping can be parameterized by the simple formula:
\begin{eqnarray}
\label{r2}\nonumber
Z_{m=1}&=&\left[Z_{m=1}\left(T_{C}\right) - Z_{m=1}\left(T_{0}\right) \right] \left(\frac{T}{T_{C}}\right)^{\Gamma}\\
&+& Z_{m=1}\left(T_{0}\right), 
\end{eqnarray}
where $Z_{m=1}\left(T_{C}\right)$ assumes the value $1.61$ for LiC$_6$ and $2.12$ for Li$_2$C$_6$. We underline that $Z_{m=1}\left(T_{C}\right)$ can be calculated on the basis of the expression: $Z_{m=1}\left(T_{C}\right)=1 + \lambda$, where the symbol $\lambda$ denotes the electron-phonon coupling constant defined as: ${\lambda\equiv 2\int^{\Omega_{\rm{max}}}_0 \alpha^2\left(\Omega\right)F\left(\Omega\right)/\Omega}$. Additionally, $Z_{m=1}\left(T_{0}\right)$ is equal to $1.6$ and to $2.01$ for LiC$_6$ and Li$_2$C$_6$, respectively.

Like in the case of the order parameter, the function $Z_{m=1}$ has the important physical interpretation. In particular, it allows to calculate the dependence of the electron effective mass ($m^{\star}_{e}$) on the temperature: $m^{\star}_{e}\simeq Z_{m=1} m_{e}$, where $m_{e}$ denotes the band electron mass.

On the basis of figure \fig{f2}, it is easily to see that the ratio $m^{\star}_{e}/m_{e}$ strongly increases with the growth of the lithium doping.

In order to determine the temperature dependence of the thermodynamic critical field and the specific heat, the free energy difference between the superconducting and normal state has been calculated:
\begin{eqnarray}
\label{r10}
\frac{\Delta F}{\rho\left(0\right)}&=&-\frac{2\pi}{\beta}\sum_{n=1}^{M}
\left(\sqrt{\omega^{2}_{n}+\Delta^{2}_{n}}- \left|\omega_{n}\right|\right)\\ \nonumber
&\times&(Z^{S}_{n}-Z^{N}_{n}\frac{\left|\omega_{n}\right|}
{\sqrt{\omega^{2}_{n}+\Delta^{2}_{n}}}),
\end{eqnarray}  
where the symbol $\rho(0)$ denotes the electron density of states at the Fermi level, and $Z^{S}_{n}$ and $Z^{N}_{n}$ represent the wave function renormalization factor for the superconducting ($S$) and normal ($N$) state.

\begin{figure}[ht]
\includegraphics[width=\columnwidth]{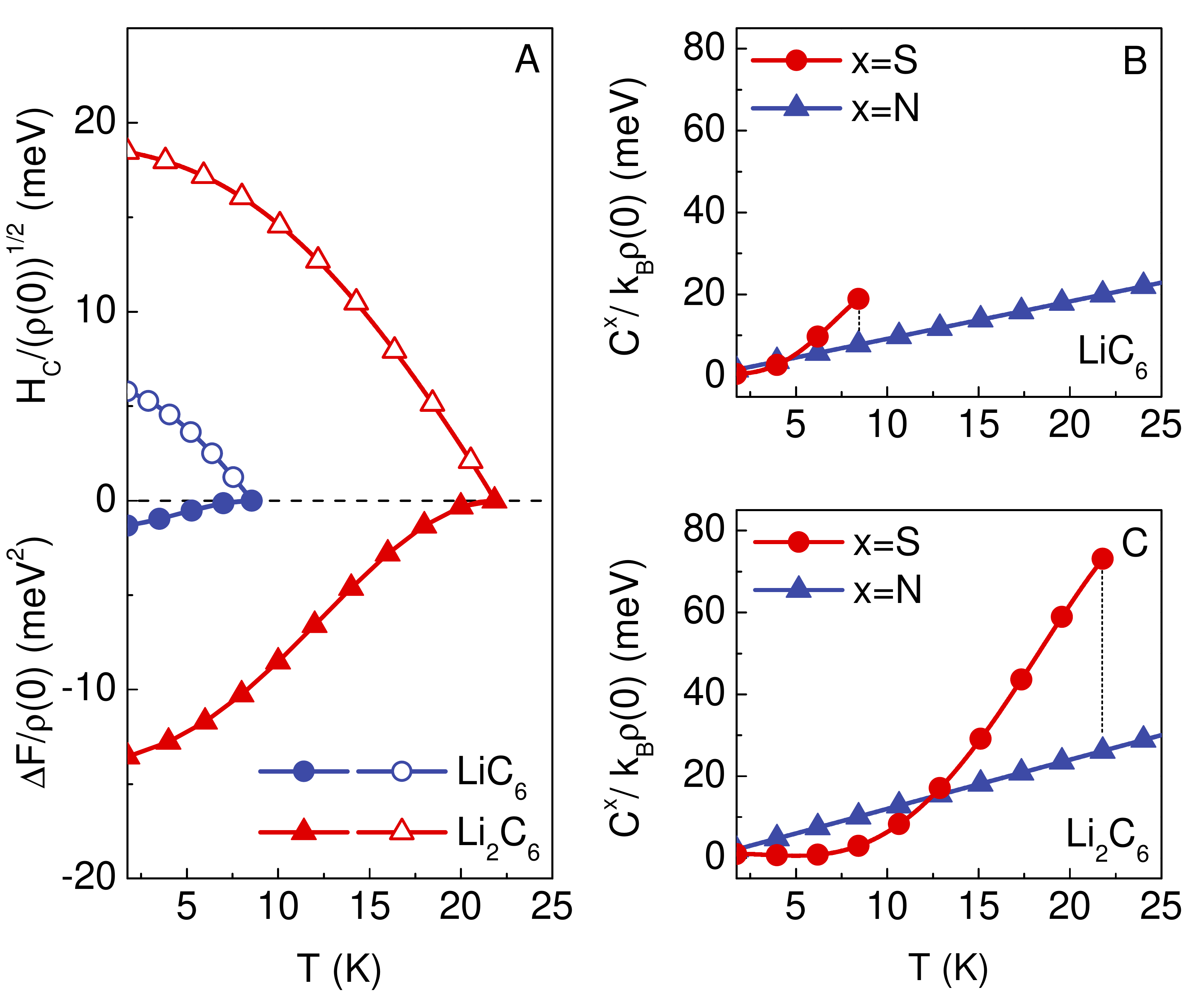}
\caption{(A) The free energy difference (lower panel) and the thermodynamic critical field (upper panel) as a function of the temperature. 
The specific heat of the superconducting and normal state as a function of the temperature for LiC$_6$ (B) and Li$_2$C$_6$ (C).}
\label{f4}
\end{figure}

The results obtained for the ratio $\Delta F/\rho\left(0\right)$ have been presented in the lower panel of Fig. \ref{f4} (A). From the physical point of view, $\Delta F/\rho\left(0\right)$ determines the thermodynamic stability of the superconducting phase. Taking into account the data plotted in Fig. \ref{f4} (A), it can be seen that the increasing lithium doping substantially strengthens the stability of the superconducting state in graphene.

The thermodynamic critical field has been given by:
\begin{equation}
\label{r11}
\frac{H_{C}}{\sqrt{\rho\left(0\right)}}=\sqrt{-8\pi\left[\Delta F/\rho\left(0\right)\right]}.
\end{equation}

In the upper panel of Fig. \ref{f4} (A), we have shown the dependence of $\frac{H_{C}}{\sqrt{\rho\left(0\right)}}$ on the temperature. The influence of the lithium doping on the critical field is also very big. 

\begin{figure*}[ht]
\includegraphics[width=\textwidth]{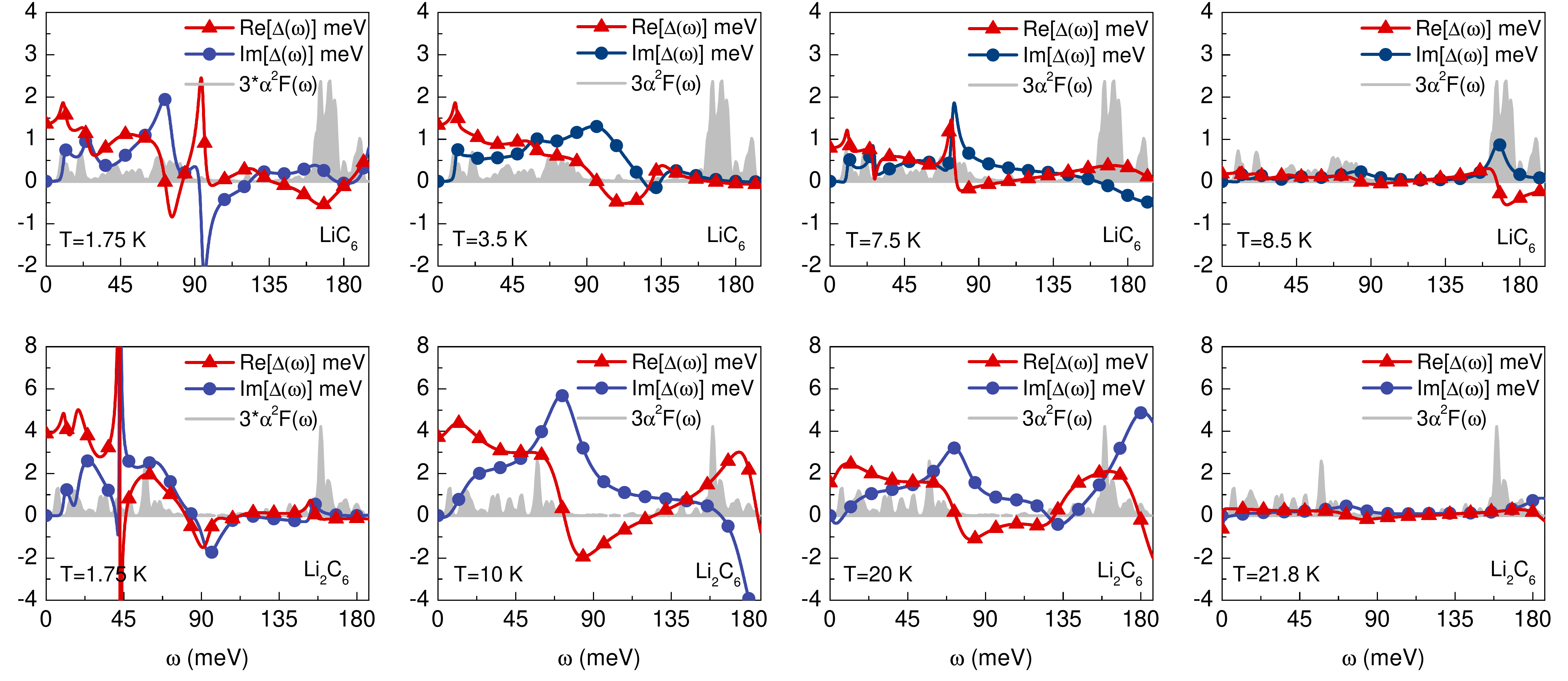}
\caption{The real and imaginary part of the order parameter on the real axis for the selected values of temperature. 
The upper row presents the data for LiC$_6$, the lower row gives the results for Li$_2$C$_6$. The rescaled Eliashberg function has been also presented.}
\label{f5}
\end{figure*}

The difference between the specific heat of the superconducting ($C^{S}$) and normal ($C^{N}$) state can be calculated by using the expression:
\begin{equation}
\label{r12}
\frac{\Delta C\left(T\right)}{k_{B}\rho\left(0\right)}=-\frac{1}{\beta}\frac{d^{2}\left[\Delta F/\rho\left(0\right)\right]}{d\left(k_{B}T\right)^{2}}.
\end{equation}
On the other hand, the normal state specific heat is given as:
\begin{equation}
\label{r13}
\frac{C^{N}\left(T\right)}{ k_{B}\rho\left(0\right)}=\frac{\gamma}{\beta}, 
\end{equation}
where the Sommerfeld constant has the form: $\gamma\equiv\frac{2}{3}\pi^{2}\left(1+\lambda\right)$.

In Fig. \ref{f4} (B) and (C), the specific heat for the normal and superconducting state has been presented. For both considered compounds, the characteristic specific heat jump at the critical temperature has been marked by the vertical line.

The results obtained for the thermodynamic critical field and the specific heats allows us to estimate the values of the corresponding characteristic dimensionless ratios \cite{bardeen1}, \cite{bardeen2}:
\begin{equation}
\label{r14}
R_{H}\equiv\frac{T_{C}C^{N}\left(T_{C}\right)}{H_{C}^{2}\left(0\right)},
\quad {\rm and} \quad
R_{C}\equiv\frac{\Delta C\left(T_{C}\right)}{C^{N}\left(T_{C}\right)}.
\end{equation}

In particular, we have: R$_{\rm H}$=0.167 and R$_{\rm C}$=1.47 for LiC$_6$, and R$_{\rm H}$=0.144 and R$_{\rm C}$=1.79 for Li$_2$C$_6$. We notice that the BCS theory predicts: R$_{\rm H}$=0.168 and R$_{\rm C}$=1.43.

In first part of the paper, we have presented our estimations of the energy gap on the basis of the Eliashberg solution on the imaginary axis. However, in order to determine the exact value of the order parameter, the solutions of the Eliashberg equations on the imaginary axis should be analytically continued on the real axis ($\omega$). For this purpose, we have numerically solved the Eliashberg equations in the mixed representation.

The obtained results allows us to estimate the physical value of the order parameter \cite{eliashberg}, \cite{carbotte}:

\begin{equation}
\label{r15}
\Delta\left(T\right)={\rm Re}\left[\Delta\left(\omega=\Delta\left(T\right),T\right)\right].
\end{equation}

In particular, the extracted values of the order parameter close to the zero Kelvin are: 
$\Delta\left(0\right)=1.37$ meV for LiC$_6$, and $\Delta\left(0\right)=3.96$ meV for Li$_2$C$_6$.

The real-axis analysis let us additionally determine the dimensionless ratio: $R_{\Delta}\equiv 2\Delta\left(0\right)/k_{B}T_{C}$. In our case, the estimated values are following: R$_{\Delta}$=3.72 for LiC$_6$, and R$_{\Delta}$=4.21 for Li$_2$C$_6$. Above results indicate that $R_{\Delta}$ for Li$_2$C$_6$ considerably exceeds the value predicted by the BCS theory ($\left[R_{\Delta}\right]_{\rm BCS}=3.53$) \cite{bardeen1}, \cite{bardeen2}.

\section{Summary}

In the present paper, we have calculated the thermodynamic properties of the lithium doped graphene superconductors LiC$_6$ and Li$_2$C$_6$.

Our analysis has predicted that the value of the critical temperature strongly rises together with the increase of the lithium doping. In particular, from $8.6$ K for LiC$_6$ to $21.8$ K for Li$_2$C$_6$.

Moreover, the calculated values of the dimensionless parameters, which describe the zero-temperature energy gap to the critical temperature, the ratio of the specific heats, as well as the ratio connected with the zero-temperature thermodynamic critical field, exceed the predictions of the BCS theory (see table \ref{tab1}). The discrepancies between our results and the BCS estimates arise due to the retardation and strong-coupling effects.

To this end, the electron effective mass also increases together with the lithium doping.

\begin{table}[ht]
\caption{\label{tab1} The values of the main thermodynamic parameters for the LiC$_6$ and Li$_2$C$_6$ monolayers.}
\begin{ruledtabular}
\begin{tabular}{c c c c c}
\\
 & $R_{\Delta}$ & $R_{H}$ & $R_{C}$ & $(m^{\star}_{e})_{\rm max}/m_{e}$\\
\\
\hline\\

LiC$_6$ & 3.72 & 0.167 & 1.47 & 1.61 \\
\\
\hline\\

Li$_2$C$_6$ & 4.21 & 0.144 & 1.79 & 2.12 \\
\\
\end{tabular}
\end{ruledtabular}
\end{table}
 
\begin{acknowledgments}

D. Szcz{\c e}{\' s}niak, would like to note that this work has been financed by the Polish National Science Center (grant DEC-2011/01/N/ST3/04492).

Additionally, we are grateful to the Cz{\c{e}}stochowa University of Technology - MSK CzestMAN for granting access to the computing infrastructure built in the project No. POIG.02.03.00-00-028/08 "PLATON - Science Services Platform".

\end{acknowledgments}
\bibliographystyle{apsrev}
\bibliography{manuscript}
\end{document}